# New Optical Gating Technique for Detection of Electric Field Waveforms with Subpicosecond Resolution

A. Muraviev[1], A. Gutin[1], G. Rupper[2], S. Rudin[2], X. Shen[1], Y. Yamaguchi[1], G. Aizin[3] and M. Shur[1]

[1]Rensselaer Polytechnic Institute, Troy, New York, USA
[2]U.S. Army Research Laboratory, Adelphi, Maryland, USA
[3]Kingsborough College, The City University of New York, Brooklyn, New York 11235, USA
*Corresponding author: Michael Shur shurm@rpi.edu

We report on the new optical gating technique used for the direct photoconductive detection of short pulses of terahertz radiation with the resolution up to 250 femtoseconds. The femtosecond optical laser pulse time delayed with respect to the THz pulse generated a large concentration of the electron hole pairs in the AlGaAs/InGaAs High Electron Mobility Transistor (HEMT) drastically increasing the conductivity on the femtosecond scale and effectively shorting the source and drain. This optical gating quenched the response of the plasma waves launched by the THz pulse and allowed us to reproduce the waveform of the THz pulse by varying the time delay between the THz and quenching optical pulses. The results are in excellent agreement with the electro-optic effect measurements and with our hydrodynamic model that predicts the ultra-fast transistor plasmonic response at the time scale much shorter than the electron transit time, in full agreement with the measured data.

## Introduction

The emergence of high resolution video stimulated interest in future wireless communications operating at frequencies of 300 GHz and above [1,2,3,4], which, in turn, stimulated research on new terahertz detectors and sources [5,6,7,8], including plasmonic devices [9,10,11,12,13], operating in both collision dominated [14], and quasi ballistic regimes [15]. Theoretical predictions [16,17] show that these plasmonic devices could respond at terahertz frequencies making them ideal for applications as modulators, detectors, and, potentially, sources for the THz applications, including wireless communications. In this paper, we report on the new optical gating technique that was used for the direct photoconductive detection of short pulses of terahertz radiation with the resolution up to 250 femtoseconds. The measurement results confirm that the transistor plasmonic response is at the subpicosecond scale in a good agreement with the predictions of our hydrodynamic model. The proposed optical gating technique could find applications for the time resolved measurements of ultra-short electric field pulses at femtosecond scale.

The proposed technique uses the quenching of the THz response with a femtosecond laser pulse. Prior to the application of the optical pulse, the plasma waves launched by the THz pulse induce a drain to source voltage. In contrast to the CW THz signal, which is proportional to the squared electric field for low intensity excitation and to the electric field for high intensity THz field [18], the instantaneous response is proportional to the THz electric field. This is confirmed by the calculations of the response as a function of the

THz pulse amplitude in the frame of the full hydrodynamic model [19], see Figure 1. The characteristic HEMT response time depends on the duration of the THz pulse, $\tau_{THz}$, the plasma wave transit time, $\tau_p$, and the momentum relaxation time, $\tau_m$.

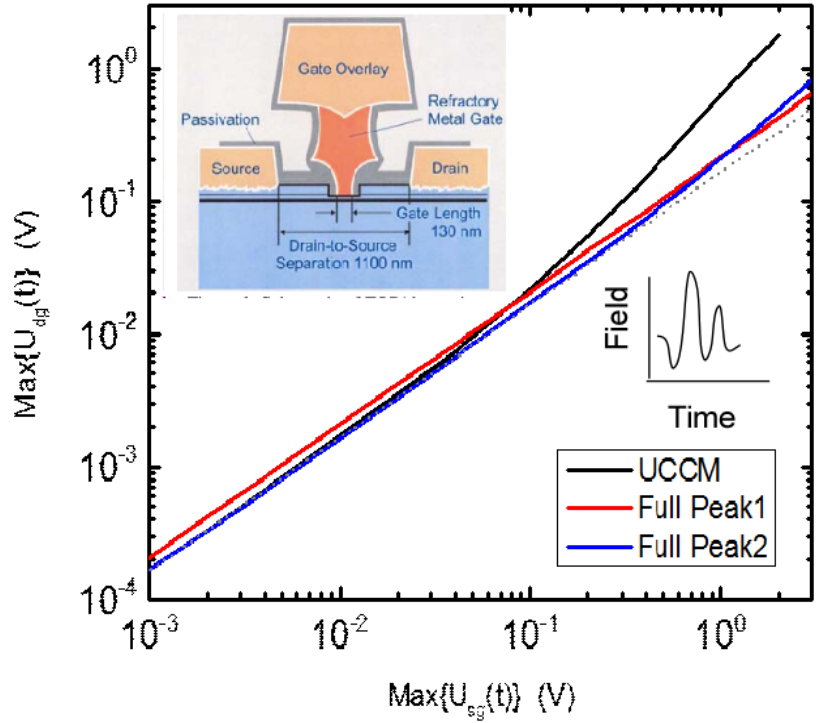

Figure 1. Computed THz response as a function of the THz pulse amplitude. The insets show the qualitative shape of the THz pulse and the device geometry[20]. The dotted line corresponds to a linear response. The device dimensions: $L_g$=130nm, $L_{sd}$=1100nm, $d$=21.97nm. The dielectric permittivity of the barrier layer $\varepsilon$=13.9.

The femtosecond laser pulse delivered to the HEMT generates a large concentration of electron-hole pairs, increases the channel conductance effectively shorting the source and drain (Figure 2).

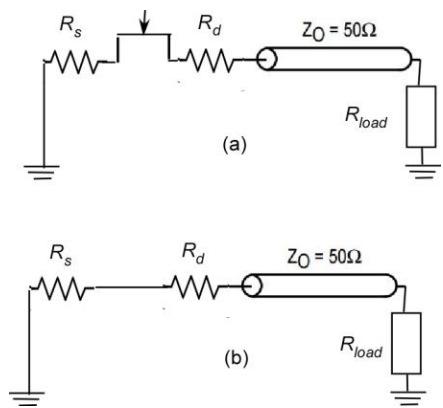

Figure 2. Equivalent circuit of the THz FET detector before (a) and after (b) laser pulse application.

The HEMT delivers a pulse to a transmission line with the duration much smaller than the characteristic transmission line propagation time. The laser pulse switches off the device response as schematically shown in Fig. 3 limiting the charge delivered to the transmission line and, hence, decreasing the amplitude of the response.

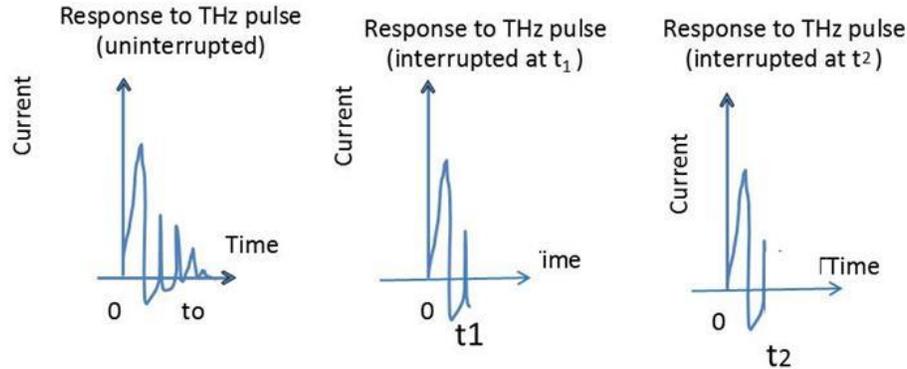

Figure 3. FET response to a picosecond THz pulse (gate-to-source transient current). (a) Non interrupted by laser pulse (b) Interrupted by laser pulse at $t = t_1$ (c) Interrupted by laser pulse at $t = t_2$.

The shape of the output pulse is determined by the transmission line and electronics readout circuit but the magnitude of the response is proportional to the drain-to-source voltage pulse Vp generated by the HEMT in response to the THz pulse.

**Experimental Technique**

For experimental study of the HEMT response on THz radiation, we used the femtosecond Tsunami/Legend laser system with the wavelength of 800 nm, pulse energy ~0.6 mJ, pulse duration of 120 fs and repetition rate of 1 kHz. A nonlinear ZnTe crystal exposed to the femtosecond laser pulses generated terahertz pulses focused on the active element of the device by a gold off-axis parabolic mirror. The amplitude of the THz electric field in the focal plane of the mirror (at the device) was 500-1000 V/cm (measured by the electro-optic technique[21, 22, 23]), and the duration of the generated THz pulse was 1-2 ps.

To scan the device response with a sub picosecond resolution in the photoconductive regime of the HEMT operation, a small fraction of the optical power (with the pulse energy of 0.15 - 5 micro Joule) was focused on the device channel with the controlled time delay with respect to the start of the THz pulse. This optical power was high enough to generate a large number of the electron hole pairs to drastically increase the conductivity on the femtosecond scale effectively shorting the source and drain. This effect quenched the response and therefore allowed us to reproduce the waveform of the THz pulse by changing the time delay between the THz pulse and quenching optical pulse.

Figure 4a shows direct photovoltaic response of the FET on a single THz pulse. The shortest obtained pulse duration of the response at 50 ps was limited by the bandwidth of

RF FET package and used acquisition electronics. Its shape is typical for the transmission line response to a short input pulse.

Figures 4b and 4c show the magnitude of the response in the photoconductive regime of the detector operation when along with the THz pulse we simultaneously irradiated the device by the optical 120 fs pulse with a variable delay time between these pulses as a function of the optical pulse delay in respect to the THz pulse.

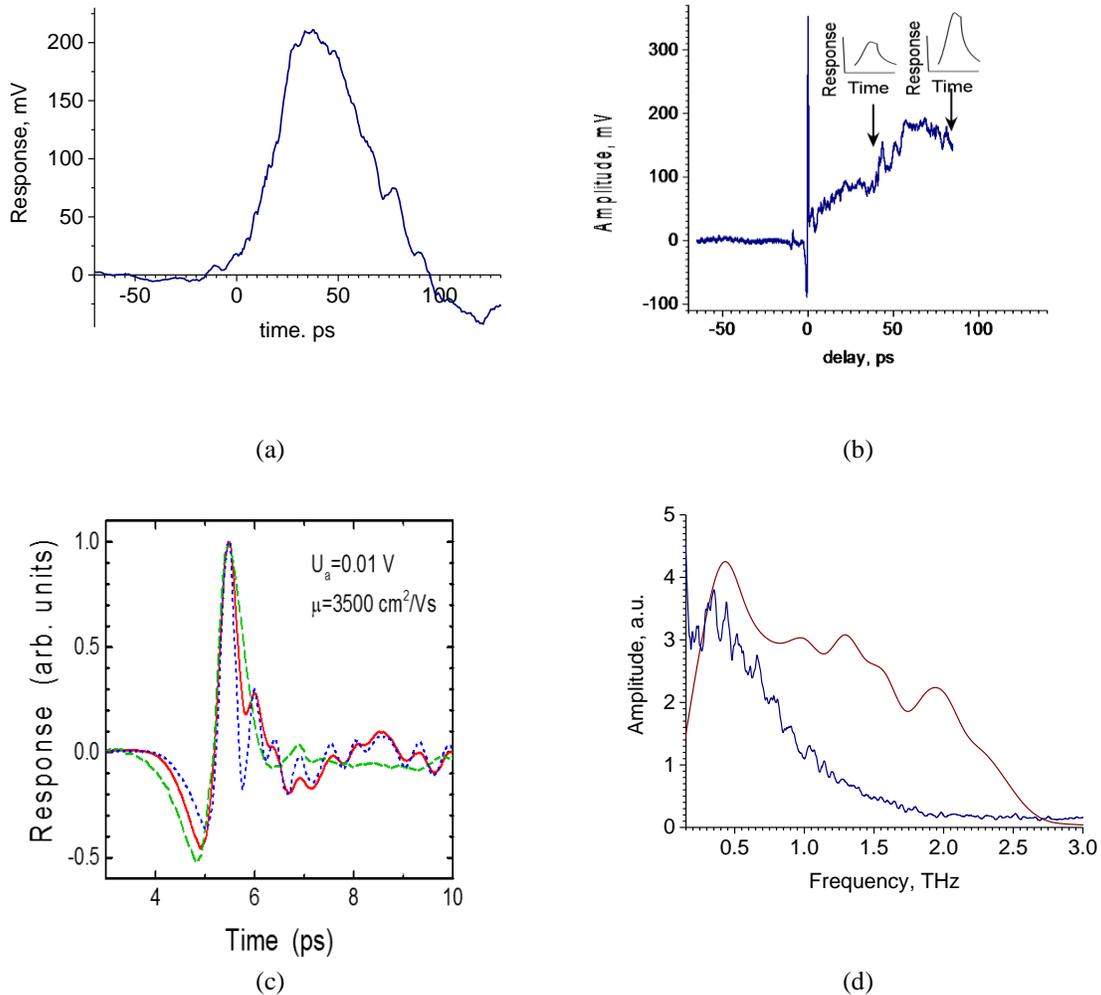

(a)    (b)

(c)    (d)

Figure 4. (a) – Photovoltaic response of the detector on a single ultra-short THz pulse. The pulse width of 50 ps (FWHM) is limited by the bandwidth of detector RF package and acquisition electronics. (b) - Dependence of the photovoltaic response amplitude in case of optical gating as a function of optical pulse delay in respect to the THz pulse. The two insets show the output pulse shapes with the varying amplitude for two different time delays. (c) - Comparison of the photoconductive response (dashed line) with electro-optic effect data (dotted line) and hydrodynamic simulations (solid line). (d) - Fourier spectra of the THz pulse electro-optic measurements (dotted line in Fig.4c) and photoconductive FET response (dashed line in Fig. 4c).

The response dependence on the time delay shown in Fig. 4b has a sharp peak at the zero delay time between the optical and THz pulses, which is caused by an increasing number of carriers during the optical pulse, and, thus, a stronger charge separation in the channel of the device induced by the THz electric field due a strong transport nonlinearies induced by this electron-hole pair generation process. The second feature is the rising tail of the THz response at large positive delays (greater than approximately 80 ps.) As seen from Figure 4a, such delays are long enough to avoid quenching before recording the maximum response of approximately 200 mV seen in both Figures 4a and 4b. Figure 4c shows the response determined from the photoconductive response as a function of the delay time on a much greater scale then Figure 5b and compares the photoconductive response measured using the novel photoconductive technique described above with the electro-optic THz detection data using ZnTe crystal and the response predicted by our hydrodynamic simulations. Figure 4d compares the Fourier spectra of the time-domain response on THz pulse, measured by the standard electro-optic technique using ZnTe nonlinear crystal (top curve)) and calculated from the optically gated HEMT response shown on Fig. 4c. The FET spectrum demonstrate narrower bandwidth in respect to ZnTe. However, the signal to noise ratio stays acceptable up to 2 – 2.5 THz. The highest frequency of THz radiation detected by this device exposed to the CW gas terahertz laser was 4.25 THz.

**Results and Discussion**

Our hydrodynamic model includes the effects of viscosity, temperature and pressure gradients, with the last one shown to be a dominant factor in the sub-threshold regime [24]. The simulation was performed solving the Poisson equation coupled with the hydrodynamic model equations. The model accounts for a gradual transition from the ungated to gated plasmons due to the non-local relation between electron density and electric potential in the channel. The results are compared with the analytical calculation for the gated region in the frame of the unified charge control model (UCCM).[25] The device modeled, has a small (4nm) gate-to-channel separation and a 130 nm gate length, so that the UCCM is expected to perform well within the gated region. The results shown in Figure 4c are in excellent agreement with the results of the electro-optic measurements and simulations in the frame of the hydrodynamic model described in [24]. This model predicted the ultra-fast transistor plasmonic response in full agreement with our measurement results. This agreement confirms the applicability of this new optical gating technique for monitoring the waveforms of short THz pulses.

**Conclusions**

We presented a new simple photoconductive technique for the measurements of ultrafast transients with a subpicosecond resolution. The results are in good agreement with a well-established electro-optic technique and with the simulations in the frame of the hydrodynamic model. They confirm a big potential of the plasmonic FET detectors for THz communications.

**Acknowledgements**. The work at RPI (M. Shur) was supported in part by the U.S. Army Research Laboratory through the Collaborative Research Alliance (CRA) for Multi-Scale Modeling of Electronic Materials (MSME).